\documentclass[prd,preprintnumbers,showpacs,12pt]{revtex4}
\usepackage[active]{srcltx}
\usepackage{epsfig}
\usepackage{graphicx}
\usepackage{bm}
\usepackage{amsmath}
\usepackage{amssymb}

\newcommand{\be}{\begin{equation}}
\newcommand{\dd}{\displaystyle}
\newcommand{\ee}{\end{equation}}
\newcommand{\bea}{\begin{eqnarray}}
\newcommand{\eea}{\end{eqnarray}}

\newcommand{\nn}{\nonumber}
\newcommand{\de}{\partial}

 \def\slash#1{\setbox0=\hbox{$#1$}#1\hskip-\wd0\dimen0=5pt\advance
       \dimen0 by-\ht0\advance\dimen0 by\dp0\lower0.5\dimen0\hbox
         to\wd0{\hss\sl/\/\hss}}

\begin{document}
\hfill{\bf ICCUB-14-049~~~\,}\vskip0.1cm

\title{Conformal symmetry for relativistic point particles}

\author{Roberto Casalbuoni}\email{casalbuoni@fi.infn.it}
\affiliation{Department of Physics and Astronomy, University of Florence and
INFN, 50019 Florence,  Italy}
\author{Joaquim Gomis} \email{gomis@ecm.ub.es}
\affiliation{
 Departament d'Estructura i Constituents de la
Mat\`eria and Institut de Ci\`encies del
Cosmos, Universitat de Barcelona, Diagonal 647, 08028 Barcelona,
Spain}

\begin{abstract}

In this paper we consider  conformal dynamics for a system of $N$  interacting relativistic massless particles. 
A detailed study is done for the case of two-particles, with a particular attention to the symmetries of the problem. In fact, we show that this analysis could be extended to the case of higher spin symmetries. Always in the two particle case a formulation in terms of bilocal fields is proposed. 
For a system of $N$ particles we consider two possible scenarios: i)  the action is invariant under any permutation of the $N$ particles. This case corresponds to completely democratic interactions with each particle interacting with all the others. The action depends on $N-1$ dimensionless coupling constants. ii) By putting the particles along a one-dimensional lattice (open or closed) with nearest neighbor interactions, one obtains a model with only two-body interactions depending on a single dimensionless coupling. This model can be easily extended to the continuum case, obtaining a conformal string-like (closed or open) system. 
\end{abstract}

\maketitle

\section{Introduction}\label{sec:0}

The idea of conformal invariance in physics is more than one hundred years old (a very nice  history of the conformal group can be found in H.A. Kastrup \cite{Kastrup:2008jn}). It started with H. Bateman that in 1908 proved the invariance of the wave equation under inversion $x^\mu\to x^\mu/x^2$. This is a discrete transformation that, as we shall see, will play a crucial role in the present paper. A few months later, Biggs himself with two papers followed by one by E. Cunningham proved that the Maxwell equations are invariant under the conformal group. The next  step was by H. Weyl in 1918, who tried  to unify the gravitational and the electromagnetic interactions making use of  conformal invariance.  This approach was strongly criticized by Einstein. As it is well known this idea led eventually to the phase invariance of the Schr\"odinger equation and to the gauge invariance of the Maxwell theory.

The conformal invariance had a revival during the sixties and the seventies in two different areas, particle physics and critical phenomena.   The interest of  scale invariance in critical phenomena raised from the works of L.P. Kadanoff \cite{Kadanoff:1966ab} and K.G. Wilson \cite{Wilson:1973jj}
In particle physics the famous SLAC experiment on deep inelastic scattering aroused wide interest in scale symmetry and its extensions. 
The relevance of conformal symmetry in field theory was outlined by Polyakov \cite{Polyakov:1970xd} and used in the Operator Product Expansion (OPE) by K.G. Wilson \cite{Wilson:1969zs}. In the context of the OPE there was a revival of the idea of bootstrap (for a review see \cite{Ferrara:1973eg}).

Since then, the attention to conformal symmetry has been always very high. In field theory we  recall the already mentioned invariance of the Maxwell equations and of the massless Dirac equation. At the classical level the scalar theory with a $\phi^4$ self-interaction and the non-abelian Yang-Mills theories are conformal invariant. In these  cases the conformal invariance is broken by anomalies, but still it plays an important role. In condensed matter and in statistical field theory  scale invariance at the critical points is a fundamental phenomenon. We also mention the importance of two-dimensional conformal symmetry in string theory and in general in two-dimensional field theories, where the conformal group, contrarily to the case of space-dimensions different from two, is infinite dimensional. As a last point we mention the AdS/CFT
correspondence
\cite{Maldacena:1997re} \cite{Gubser:1998bc}  \cite{Witten:1998qj}
which allows to define in a non-perturbative way M/string theory in terms of a  (superconformal) quantum field theory in flat space-time. 
This idea has opened the possibility to study strongly coupled field theories in terms of gravitational theories.

 It should be  underlined that one of the main reasons that makes conformal theories so attractive is that they do not depend  on any dimensionfull  coupling constant.

More recently, conformal symmetry has become an important tool in the analysis of higher spin theories \cite{Vasiliev:1990en,Vasiliev:2003ev}, for a recent review see 
\cite{Didenko:2014dwa,Vasiliev:2014vwa}. About this point  it is interesting to notice that the higher spin symmetries of Vasiliev theory appear in the free massless Klein Gordon equation \cite{Eastwood:2002su}. At particle level these symmetries are all the symmetries of the action of a relativistic massless particle and they generalize the well know conformal symmetries of this action.

In this paper we present an application of conformal  invariance  to classical interacting relativistic  particles. 
First of all, this problem, at the best of our knowledge, has been considered only in the non-relativistic one-dimensional case. Examples are  the Calogero-Moser rational model \cite{Calogero:1970nt,Moser:1975qp,Peremelov:1983aa}, describing $N$ interacting particles via two body interactions. This model is very important  in the context of integrable models.
For an extension to the supersymmetric  case see, for example, \cite{Freedman:1990gd}. The other example, always in one dimension, is in reference  
\cite{de Alfaro:1976je}  where only one degree of freedom is considered but it contains a deep analysis of the role of the conformal group. For   the superconformal case see \cite{Fubini:1984hf}.

The second point is that, after quantization, this theory  is naturally connected with  non-local  field theories  appearing  in the context of higher spin theories, see for example  \cite{Sorokin:2004ie}. As we mentioned previously, the underlying physics of the latter theories is connected with massless free particles. Since  we are considering interacting massless particles, it would be worth to try to understand the possible connections.

This paper is organized as follows: after the Introduction, in Section 2
  we make use of the dilatation and translation invariance to show that a free massless particle cannot be described in configuration space proving that the lagrangian  vanishes identically. Therefore,  in Section 3 we introduce lagrange multipliers (or einbeins) in order to impose the mass zero condition. This is the description that we will use throughout all this paper. Furthermore, using conformal invariance we write down an action for two relativistic massless particles  in a $D$ dimensional space-time, with $D\not = 2$. In the case of two particles we  further show that it is indeed possible to write down a lagrangian using only the coordinate space. It turns out that this lagrangian vanishes identically  when turning off the coupling constant describing the interaction, as expected from the previous considerations. In this Section we show also that there is a constraint in phase space involving the product of the momenta squared of the two particles. 
 
 Section 4 is dedicated to the Hamiltonian analysis of the model. It turns out that there are two primary constraints and two secondary ones. We show that out of these four constraints two  are first class and two  second class. Eliminating the second class constraints through the use of the Dirac brackets  we  recover, in the reduced phase space, the constraint found previously in the lagrangian analysis. 

By construction, our model is explicitly invariant under the  conformal group acting upon the coordinates of the two particles, but it is interesting to study in an explicit way the Killing vectors of the model. This is done in Section 5, where we show that  both the Killing vectors associated to the two particles  satisfy the conformal Killing equations with independent infinitesimal parameters. However, due to the interaction, the two vectors must satisfy a further condition requiring that the infinitesimal parameters of the two Killing vectors coincide. This implies an explicit  breaking of the symmetry group of the free case $SO(D,2)_1\otimes SO(D,2)_2$  to the diagonal subgroup $SO(D,2)$ .

The previous study is  preliminary to what we do in Section 6, where we consider higher order Killing tensors. This means to take powers of the generator defined in the previous Section. Obviously these powers are  constant of motion, but there is some interest from the point of view of higher spin symmetries to study conformal Killing tensors \cite{Carter:1968rr}, see also the more recent papers \cite{Cariglia:2014dfa,Cariglia:2014dwa}.
 In the case of free massless  relativistic particles these symmetries are  the enveloping algebra of 
the relativistic conformal group \cite{Eastwood:2002su,Segal:2002gd}.
 We study in particular the case of a Killing tensor of rank 2, deriving the conditions that must be satisfied to provide the required invariance. The equations we get are obviously satisfied when the Killing tensor is realized as the product of two Killing vectors, so the interest is to look for non factorized solutions. Is should also be noticed that in the case of higher spin the interest is in Killing vectors corresponding to a single space-time variable, whereas in our case they depend on two space-time variables (corresponding to the fact that we are studying a two-particle system).

In Section 7 we construct a bilocal field theory, involving two bilocal fields, such to incorporate the constraint
found in Section 2. This is also an interesting point since bilocal fields are naturally connected with higher spin symmetries.

In Section 8 we extend the two-particle model to $N$ massless particles interacting in a conformal invariant way. Here various possibilities open up according to the kind of symmetry we require under the exchange of  the $N$-particles. In particular we will examine two models, in the first one we assume invariance with respect to any permutation among the $N$ particles. This entails a completely democratic model in which each particle interact with all the others. The model depends on $N-1$ dimensionless coupling constants.
We show that this model  has the remarkable property of asymptotic separability. This should be understood in the following way: if we divide the $N$ particles in two clusters one made up with $n$ and the other with $m$ particles, and we send to infinity all the distances among the particles of the first cluster and the particles of the second cluster, the original lagrangian goes into the sum of two lagrangians of the same kind of the original one. In the second model considered here, we associate the particle labels to the sites of a one-dimensional lattice, assuming nearest neighbor interactions. Therefore only two-body interactions are involved and the model is defined by a single dimensionless coupling. The asymptotic separability holds also in this case.
There are not symmetries related to the exchange of  particles. However, for a closed lattice there is  a symmetry under discrete translations. 

In Section 9 we draw some conclusions and give an outlook for further problems to be studied.

\section{Conformal invariance in   particle coordinates}\label{sec:1}

We will discuss the requirements coming from conformal invariance on the lagrangian of classical relativistic point-particles. Let us start with one-particle. We will prove that a conformal invariant lagrangian for a single relativistic particle vanishes identically. Actually it will be enough  to assume that the action is parametrization invariant and that it depends only on the coordinates of the particle. 
The generator of dilatations for a single particle is given by 
\be
D=x_\mu p^\mu=-x_\mu\frac{\de L}{\de \dot x_\mu}.\ee
Notice  the minus sign in the definition of the canonical momentum. This follows from our choice of a mostly minus metric $g_{\mu\nu}=(+,-,-,\cdots, -)$ in a $D$ dimensional space-time.
We require $D$ in the previous equation to be a constant of motion and, furthermore, that the Lagrangian is homogeneous of first degree in the time parameter. It follows
\be
0=\frac{dD}{d\tau}=-\dot x_\mu\frac{\de L}{\de \dot x_\mu}-x_\mu\frac{\de L}{\de x_\mu}=-L,\ee
where we have used the Lagrange equations of motion 
\be
\frac{d}{d\tau}\frac{\de L}{\de \dot x_\mu}=\frac{\de L}{\de x_\mu}=0\ee
and the invariance under translations.
Therefore, the only solution for $D$ to be a constant in time, is that the lagrangian vanishes. It is obvious that this result applies to the case of $N$ non interacting particles (under the same assumptions).
This is an important point, since, if we want to consider a conformal invariant theory for a given number of particles, we cannot describe the free case using only space-time variables. In fact, as it is well known, a massless particle  is  described using an einbein variable defined on the world line (in practice a Lagrange multiplier). Therefore, this is the description that we will adopt, although for more than one  particle a conformal invariant lagrangian depending only on the coordinates can be constructed. On the other hand this latter formulation is such that turning off the interaction, the lagrangian vanishes identically, as it should be clear from the previous discussion.

\section{Formulation with the einbeins}\label{sec:2}

The lagrangian for a single free massless particle can be obtained through the use of an  einbein $e$:
\be
S= -\int d\tau \frac{\dot x^2}{2e}\label{eq:2.1},\ee
from which varying with respect to the einbein we get the equation $\dot x^2=0$ and evaluating the momentum $p_\mu =\dot x_\mu/e$ we obtain $p^2=0$.  The minus sign in front of the action is a consequence of our choice of the space-time metric. Requiring that the einbein transforms as a time derivative, this action is invariant under  reparametrization. It is also invariant under  Poincar\'e transformations. As for dilatations, we require:
\be
x_\mu\to \lambda x_\mu,~~~e\to\lambda^2 e.\ee
Furthermore, we recall that a special   conformal transformation can be obtained  through the following series of operations:  (inversion)$\otimes$(translation)$\otimes$(inversion), therefore, to impose the conformal symmetry  it is enough to require the invariance under inversion
 \be
 x^\mu\to \frac{x^\mu}{x^2}.\ee
 The transformation properties  of  $\dot x^2$ is
 \be
 \dot x^2\to \frac{\dot x^2}{x^4},\ee
from which it follows
 \be
 e\to \frac e{x^4}.\ee
Summarizing, the action (\ref{eq:2.1}) is  invariant  under conformal and reparametrization transformations.
 
 Now let us discuss the case of two particles. We start at the free level with two massless particles
 \be
 S_{free}= -\int d\tau \left(\frac{\dot x_1^2}{2e_1} + \frac{\dot x_2^2}{2e_2}\right).\ee
 In order to construct an interaction term depending on the relative coordinate
 \be
 r_\mu=x_{1\mu}-x_{2\mu},\ee
 we notice that under inversion
 \be
 r^2\to \frac{r^2}{x_1^2 x_2^2}.\ee
 Therefore  a conformal invariant action for two  relativistic particles is given by
\be
S= -\int d\tau \left(\frac{\dot x_1^2}{2e_1} + \frac{\dot x_2^2}{2e_2}+\frac{\alpha^2}4\frac{\sqrt{e_1e_2}}{r^2}\right).\label{eq:3.6}\ee
 The variation with respect to the einbeins gives rise to the following equations
 \bea
\frac{\de L}{\de e_1}&=&\frac{\dot x_1^2}{2e_1^2}-\frac{\alpha^2}8\sqrt{\frac{e_2}{e_1}}\frac 1{r^2}=0,\cr
\frac{\de L}{\de e_2}&=&\frac{\dot x_2^2}{2e_2^2}-\frac{\alpha^2}8\sqrt{\frac{e_1}{e_2}}\frac 1{r^2}=0.\label{eq:3.10}\eea
Resolving these two equations in the einbeins one finds
\be
\frac 1{e_1}=\frac{\alpha}{2\dot x_1^2}\left(\frac{\dot x_1^2\dot x_2^2}{r^4}\right)^{1/4},~~~
\frac 1{e_2}=\frac{\alpha}{2\dot x_2^2}\left(\frac{\dot x_1^2\dot x_2^2}{ r^4}\right)^{1/4},\ee
 with $\alpha\ge 0$.  In extracting the square root we have chosen the minus sign, in order to have the time component of the canonical momenta with the same sign of the time derivative of the coordinate times, $x_i^0$. Substituting inside the action (\ref{eq:3.6}) we find
 \be
 S=-\alpha\int d\tau \left(\frac{\dot x_1^2\dot x_2^2}{ r^4}\right)^{1/4}.\label{eq:3.12}\ee 
 As we have discussed previously the conformal invariant action for two particles in configuration space vanishes when the interaction is turned off.
 
 Evaluating the momenta from (\ref{eq:3.6}) we get
 \be
 p_i^\mu=-\frac{\de L}{\de \dot x_{i\mu}}=
 \frac{\dot x_i^\mu}{e_i}.\label{eq:3.13}\ee
 The equations (\ref{eq:3.10}) can be expressed in terms of the momenta obtaining
 \be
 p_1^2 -\frac{\alpha^2}4\sqrt{\frac{e_2}{e_1}}\frac 1 {r^2}=0,~~~~ p_2^2 -\frac{\alpha^2}4{\sqrt{\frac{e_1}{e_2}}}\frac 1 {r^2}=0. \label{eq:3.14a}\ee
 Finally, eliminating the ratio $e_1/e_2$ from these two equations we get a constraint among momenta and coordinates
 \be
 p_1^2p_2^2-\frac{\alpha^4}{16 r^4}=0.\label{eq:3.14}\ee
 This relation can also be obtained as a primary constraint 
 from the action (\ref{eq:3.12}).
 
 Notice that we have started with a flat metrics $g_{\mu\nu}$, but we could have started with a conformal metrics as well, $g_{\mu\nu}\to  \exp(2\gamma(x))g_{\mu\nu}$. In fact, in the formulation (\ref{eq:3.6}), the conformal factor can be absorbed into the definition of the einbeins, whereas the formulation (\ref{eq:3.12}) is explicitly scale invariant.

 \section{Hamiltonian Analysis}
 
 In our notations the Poisson brackets among coordinates and momenta are:
\be\{x^\mu, p^\nu\}=-g^{\mu\nu},~~~~\{e_i,\pi_j\}=\delta_{ij}.
\ee
 Once again, the sign in the first Poisson bracket is fixed by our choice of the mostly minus metric.
Notice also that there are two primary constraints
\be\pi_i=\frac{\de L}{\de \dot e_i}=0,\ee
therefore the canonical hamiltonian results to be
\be
H_C=-p_1\dot x_1-p_2\dot x_2-L=-\frac {e_1}2 p_1^2-\frac {e_2}2 p_2^2+\frac{\alpha^2}4\frac{\sqrt{e_1e_2}}{r^2}.
\ee
 Following Dirac we define the Dirac Hamiltonian, $H_D$, adding an arbitrary combination of the primary constraints $\pi_i=0$, in terms of two arbitrary functions $\lambda_i$,
 \be
 H_D=H_C+\lambda_1\pi_1+\lambda_2\pi_2.\ee
 Requiring the stability of the primary constraints we get two secondary constraints
\bea
&&\{\pi_1,H_D\}=\frac 12\left(p_1^2-\frac{\alpha^2}4\sqrt{\frac{e_2}{e_1}}\frac 1 {r^2}\right)\equiv \phi_1,\cr
&&\{\pi_2,H_D\}=\frac 12\left(p_2^2-\frac{\alpha^2}4\sqrt{\frac{e_1}{e_2}}\frac 1 {r^2}\right)\equiv\phi_2.
\eea
Notice that these two constraints are the same as the ones in (\ref{eq:3.14a}).
 Then, we have to consider the stability of the secondary constraints $\phi_i$, obtaining
\bea
&&\{\phi_1,H_D\}=-\frac{\alpha^2}{4r^4}\left(\sqrt{e_1e_2} p_1\cdot r+\sqrt{\frac{e_2^3}{e_1}} p_2\cdot r\right)+\frac{\alpha^2}{16 r^2}\left(\lambda_1\sqrt{\frac{e_2}{e_1^3}}-\lambda_2\frac1{\sqrt{e_1e_2}}\right),\cr
&&\{\phi_2,H_D\}=+\frac{\alpha^2}{4r^4}\left(\sqrt{\frac{e_1^3}{e_2}} p_1\cdot r+\sqrt{e_1e_2} p_2\cdot r\right)+\frac{\alpha^2}{16 r^2}\left(\lambda_2\sqrt{\frac{e_1}{e_2^3}}-\lambda_1\frac1{\sqrt{e_1e_2}}\right).\label{eq:4.5}\eea
These two constraints are not independent. In fact, the second equation can be obtained from the first one multiplying by $-e_1/e_2$. It follows that  the stability of the secondary constraints can be attained by eliminating one of the two parameters $\lambda_i$, for instance, evaluating $\lambda_1$ from the first equation (\ref{eq:4.5}). We find
\be
\lambda_1=\frac{e_1}{e_2}\lambda_2+\frac 4{r^2}\left(e_1^2 p_1\cdot r+ e_1 e_2 p_2\cdot r\right).\ee
Correspondingly the Dirac hamiltonian becomes
\be
H_D=H_C+\frac 4{r^2}e_1\left(e_1 p_1\cdot r+ e_2 p_2\cdot r\right)\pi_1+\lambda_2\left(\pi_2+\frac{e_1}{e_2}\pi_1\right).\ee
It is  convenient to redefine $\lambda_2=\tilde\lambda_2 e_2$, then 
\be
H_D=H_C+\tilde\lambda_2\left(e_1\pi_1+{e_2}\pi_2\right)+C\pi_1,\ee
with
\be
C=\frac 4{r^2}e_1\left(e_1 p_1\cdot r+ e_2 p_2\cdot r\right).\ee
This expression suggests that the coefficient of $\lambda_2$ is a first class constraint. This can be verified by evaluating  its Poisson bracket with $H_D$
\be
\{e_1\pi_1+e_2\pi_2,H_D\}=e_1\phi_1+e_2\phi_2-C\pi_1.\ee
This shows that the constraint $e_1\pi_1+e_2\pi_2$ is weakly stable. Then it is a simple algebra to prove that the two constraints
\be
e_1\pi_1+e_2\pi_2,~~~e_1\phi_1+e_2\pi_2 -C\pi_1\ee
are weakly first class, that is their Poisson brackets with  the other constraints $\pi_1$, $\pi_2$, $\phi_1$ $\phi_2$ are proportional to one of these constraints. In conclusion, the four constraints can be divided as follows:
\bea
&&{\rm{first~class}}~~~~e_1\pi_1+e_2\pi_2,~~~~e_1\phi_1+e_2\phi_2-C\pi_1,\cr
&&{\rm{second~class}}~~~~\pi_1,~~~\phi_1.\eea
Then, introducing  the Dirac parentheses one can put $\pi_1$ and $\phi_1$ strongly to zero. In this way 
$\pi_2$ and $\phi_2$ turn out to be strongly first class.

The matrix of the second class constraints, $\chi_{ij}, ~i,j=1,2$, is quite simple
\be
\chi=\left(\begin{array}{cc}0 & D \\ -D & 0\end{array}\right),~~~\chi^{-1}=\left(\begin{array}{cc}0 &-1/D \\ 1/D & 0\end{array}\right),\ee
where
\be
D=\{\phi_1,\pi_1\}=\frac {\alpha^2}{16} \sqrt{\frac{e_2}{e_1^3}}\frac 1{r^2}.\ee

The Dirac brackets among any two dynamical variables are given by
\be
\{O_1,O_2\}^*=\{O_1,O_2\}+\frac 1 D\left[\{O_1,\phi_1\}\{\pi_1,O_2\}-\{O_1,\pi_1\}\{\phi_1,O_2\}\right].\ee

In the reduced space using the second class constraints and the Dirac brackets the first  class constraints become
 \be
 \pi_2=0,\quad \phi_2|_{\phi_1=0}=0, \quad  {\rm{implying}} \quad \left(p_1^2 p_2^2-\frac 1{16}\frac {\alpha^4}{r^4}\right)=0, 
 \ee
 and we recover eq. (\ref{eq:3.14}) that was obtained previously by solving the equations for the einbeins.

Since among of the two first class constraints we have one that is primary,  it is known that we should have one  gauge transformation . The generator of this gauge transformation can be constructed from a well know algorithm, 
 see for example 
 \cite{Castellani:1981us,Kamimura:1982cs,Gracia:1988xp,Henneaux:1990au,Gomis:1989vy,Lusanna:1991je}. 
 The generator $G$, which is a constant of motion, is given by
 \be
G=\sum_{i=1}^2\left(\frac{d}{d\tau}(\epsilon e_i)\pi_i-(\epsilon e_i) \phi_i\right),
\ee
where $\epsilon(\tau)$ is an arbitrary function of the global parameter that parametrize the two world lines.

The transformation generated by $G$ is 
\be
\delta e_i=\frac{d}{d\tau}(\epsilon e_i),\quad \delta x^\mu_i=\epsilon \dot x^\mu_{i},
\ee
it is the  global world line diffeomorphism (Diff). Note that the interaction breaks the
individual Diff invariance of the two world lines.

 \section{Analysis of the rigid symmetries}

We have constructed our lagrangian requiring conformal invariance, that is  invariance under the group $SO(D,2)$. Of course, when we consider  two free massless particles,  the invariance group is larger, namely it is the direct product of two conformal groups $SO(D,2)_1\otimes SO(D,2)_2$ acting  on the variables of the particles 1 and 2 respectively. When the interaction is  introduced, the invariance is broken explicitly to the diagonal subgroup. 

It is interesting to analyze these symmetries by looking at the conditions the Killing vectors must satisfy in order our lagrangian is invariant under the symmetries generated by  generic Killing vectors
\be
G=\sum_{i=1}^2\xi_{i\mu}(x_1,x_2)p_i^\mu.\label{eq:5.1}\ee

In this Section we will make use of the lagrangian $L$, given in (\ref{eq:3.12}), in terms of which we have 
\be
p_{i\mu}=-\frac{\de L}{\de \dot x_i^\mu}=-\frac 12 \frac{\dot x_{i\mu}}{\dot x_i^2}L
\label{eq:5.2}\ee
and the equations of motion
\be
\dot p_{1\mu}=-\frac{\de L}{\de x_1^\mu}=+\frac 12 \frac{r_\mu}{r^2}L,~~~
\dot p_{2\mu}=-\frac{\de L}{\de x_2^\mu}=-\frac 12 \frac{r_\mu}{r^2}L.\label{eq:5.3}\ee

It is clear that the result will be that the Killing vectors are those of the conformal group but, the equations we will find here will be important for the analysis of the Killing tensors we will do in the next Section.  This analysis is relevant for  the higher spin symmetries that have been recently considered in the literature 
\cite{Eastwood:2002su} \cite{Segal:2002gd}.
 
 By taking the time derivative of $G$, using the expression of the momenta  given in eq. (\ref{eq:5.2}) and the Lagrange equations of motion (\ref{eq:5.3}), we obtain
 \be\label{dotg}
\dot G= -\frac 12\sum_{i,j=1}^2\left(\de_{j\mu}\xi_{i\nu}(x_1,x_2)\right)\frac{\dot x_j^\mu \dot x_i^\nu}{\dot x_i^2}L+
{(\xi_{1\mu}(x_1,x_2)-\xi_{2\mu}(x_1,x_2))r^\mu}\frac L{r^2}=0,\ee
where
$
\de_{i\mu}={\de}/{\de x^{i\mu}}$. Notice that the first term of this equation is not symmetric in $j$ and $i$. 

A necessary condition to have a solution is
 \be\label{killing1}
\de_{j\mu}\xi_{i\nu}(x_1,x_2)=0,\quad j\neq i,
\ee
which implies that $\xi_{i\nu}=\xi_{i\nu}(x_i)$. If we use this information in (\ref{dotg})
 \be\label{dotg1}
-\frac 12\sum_{i=j}^2\left(\de_{i\mu}\xi_{i\nu}(x_i)\right)\frac{\dot x_i^\mu \dot x_i^\nu}{\dot x_i^2}L+
{(\xi_{1\mu}(x_1,x_2)-\xi_{2\mu}(x_1,x_2))r^\mu}\frac L{r^2}=0.\ee
 Now the first term is symmetric in $\mu, \nu$. The solution of this equation is
\be\frac 12 (\de_{i\mu}\xi_{i\nu}(x_i)+\de_{i\nu}\xi_{i\mu}(x_i))=g_{\mu\nu}\lambda_{(i)}(x_i),~~~i=1,2\label{killing2},\ee
\be
\frac 12 \sum_{i=1}^2\,\lambda_{(i)}=
{(\xi_{1\mu}-\xi_{2\mu})r^\mu}\frac 1{r^2}.\label{eq:4.2}\ee
By contracting together the indices $\mu,\nu$ in (\ref{killing2})  we find
\be
\lambda_{(i)}=\frac 1D \,\de^\rho_i\xi_{i\rho}.\label{eq:5.9}
\ee

The two equations 
(\ref{killing2}) 
tell us that  $\xi_1^\mu$ and $\xi_2^\mu$ are the Killing vectors of  two conformal groups $SO(D,2)_i$ acting on the two variables $x_1$ and $x_2$ respectively. This is the symmetry group of two massless non-interacting particles. However, it is easily proved that the second condition (\ref{eq:4.2}) is satisfied if and only if the infinitesimal parameters defining the two Killing vectors are identical. Therefore the symmetry $SO(D,2)_1\otimes SO(D,2)_2$ is broken down to the diagonal subgroup $SO(D,2)$  due to the interaction between the two particles.
 
 \section{Higher spin symmetries}
 
 In the previous Section we have shown that the quantity $G$ (see  (\ref{eq:5.1})) is a constant of motion, if the parameters defining the two conformal Killing vectors, corresponding to the two particles, are the same. It is a trivial observation that the power $G^n$ is also a constant of motion
 \be
 \frac{dG^n}{d\tau}=0.\ee
 The explicit expression for $G^n$ is
 \be
G^n=\sum_{i_1,i_2,\cdots,i_n=1}^2 \xi_{i_1\mu_1} \xi_{i_2\mu_2}\xi_{i_n\mu_n}p_{i_1}^{\mu_1}p_{i_2}^{\mu_2}\cdots p_{i_n}^{\mu_n}. \label{eq:6.1}\ee
  This defines a tensor of rank $n$  constructed in terms of the $n$ conformal Killing vectors in $n$ variables $x_1, x_2,\cdots, x_n$.

In principle, one could try to generalize the expression (\ref{eq:6.1}) to a generic Killing tensor \cite{Carter:1968rr}\be
 G'=\sum_{i_1,i_2,\cdots,i_n=1}^2 \xi_{i_1i_2\cdots i_n}^{\mu_1\mu_2\cdots\mu_n }p_{i_1\mu_1}p_{i_2\mu_2}\cdots p_{i_n\mu_n}. \ee
Notice that the tensor $ \xi_{i_1i_2\cdots i_n}^{\mu_1\mu_2\cdots\mu_n }$ may depend on the  variables $x_1, x_2,\cdots, x_n$. Requiring  $G'$ to be a conserved quantity one gets
\bea
0&=&\sum_{k,i_1,i_2,\cdots,i_n=1}^2\de_{k}^\mu \xi_{i_1i_2\cdots i_n}^{\mu_1\mu_2\cdots\mu_n }p_{i_1\mu_1}p_{i_2\mu_2}\cdots p_{i_n\mu_n}\dot x_{k\mu}+\nn\\ &+&\sum_{j=1}^n\sum_{i_1,i_2\cdots,i_n=1}^2
 \xi_{i_1i_2\cdots i_n}^{\mu_1\mu_2\cdots\mu_n }p_{i_1\mu_1}p_{i_{j-1}\mu_{j-1}}\left( (-1)^{i_{j}-1}\frac{r_\mu}{r^2} L\right) p_{i_{j+1}\mu_{j+1}}\cdots p_{i_n\mu_n},
\eea
where we have used the equations (\ref{eq:5.3}) in the form
\be
\dot p_{i\mu}=(-1)^{i-1}\frac{r_\mu}{r^2}L.\ee
Then, using (\ref{eq:5.2})
\be
p_{i\mu}=-\frac 1 2\frac{\dot x_{i\mu}}{\dot x_i^2}L,\ee
\bea
&&0=\frac 12\sum_{k,i_{1},\cdots,i_{n}=1}^2\de_{k}^\mu \xi_{i_1i_2\cdots i_n}^{\mu_1\mu_2\cdots\mu_n }\frac{\dot x_{i_1\mu_1}\dot x_{i_2\mu_2}\cdots \dot x_{i_n\mu_n}\dot x_{k\mu}}{\dot x_{i_{1}}^2x_{i_{2}}^2\cdots x_{i_{n}}^2}\nn\\&& -\sum_{j=1}^n\sum_{i_1,i_2\cdots,i_n=1}^2
 \xi_{i_1i_2\cdots i_n}^{\mu_1\mu_2\cdots\mu_n }\frac{\dot x_{i_1\mu_1}\dot x_{i_{j-1}\mu_{j-1}}}{\dot x_{i_{1}}^2x_{i_{2}}^2\cdots x_{i_{j-1}}^2}\left( (-1)^{i_{j}-1}\frac{r_\mu}{r^2} \right) \frac{\dot x_{i_{j+1}\mu_{j+1}}\cdots \dot x_{i_n\mu_n}}{\dot x_{i_{j+1}}^2\dot x_{i_n}^2}.\eea
Proceeding  as in the previous Section, one obtains equations for the tensor  $ \xi_{i_1i_2\cdots i_n}^{\mu_1\mu_2\cdots\mu_n }$ independent on $\dot x_i$. We know that these equations are satisfied when the tensor factorizes in $n$ conformal Killing vectors. 

An interesting question remains open: is the factorized case the only solution to the previous equations?

Let us study in detail the case of  $n=2$.
We have
\be\label{gdotn2}
\sum_{ijk=1}^2\left(\frac 1 4\de_k^\rho\xi_{ij}^{\mu\nu}\frac{\dot x_{k\rho}\dot x_{i\mu} \dot x_{j\nu}}{\dot x_i^2\dot x_j^2}\right)- \sum_{ij=1}^2\left(\xi_{ij}^{\mu\nu}(-1)^{i-1}\frac {\dot x_{j\nu} r_\mu}{\dot x_j^2r^2}\right)=0.\ee
Notice that this equation it is not symmetric in $k, i, j$. Proceeding as in the previous section the
necessary condition to have a solution  is
\be
\de_k^\rho\xi_{ij}^{\mu\nu}=0,\quad k\neq i,k \neq j,
\ee
using this condition we have 
\bea
&&\sum_{ k=i,i\neq j}^2 2 \left(\frac 1 4\de_i^\rho\xi_{ij}^{\mu\nu}\frac{\dot x_{i\rho}\dot x_{i\mu} \dot x_{j\nu}}{\dot x_i^2\dot x_j^2}\right) +\sum_{ k=i= j}^2\left(\frac 1 4\de_i^\rho\xi_{ii}^{\mu\nu}\frac{\dot x_{i\rho}\dot x_{i\mu} \dot x_{i\nu}}{\dot x_i^2\dot x_j^2}\right)-
\nn\\
&&- \sum_{i\neq j}^2\left(\xi_{ij}^{\mu\nu}(-1)^{i+1}\frac {\dot x_{j\nu} r_\mu}{\dot x_j^2r^2}\right)
-\sum_{i=j}^2\left(\xi_{ii}^{\mu\nu}(-1)^{i+1}\frac {\dot x_{i\nu} r_\mu}{\dot x_i^2r^2}\right)
=0.\eea
We get a solution requiring
\be
\frac12(\de_i^\rho\xi_{ij}^{\mu\nu}+\de_i^\mu\xi_{ij}^{\rho\nu})=g^{\rho\mu}W_{(ij)}^\nu,\quad i\neq j,\label{eq:6.11}\ee
\be
\de_i^\rho\xi_{ii}^{\mu\nu}+\de_i^\mu\xi_{ii}^{\rho\nu}+\de_i^\nu\xi_{ii}^{\mu\rho}=
g^{\rho\mu}V_{(i)}^\nu+g^{\mu\nu}V_{(i)}^\rho+g^{\nu\rho}V_{(i)}^\mu,
\label{eq:6.12}\ee
\be
 \sum_{i=1}^2\left(2\frac 14 W_{(ij)}^\nu
- \xi_{ij}^{\mu\nu}(-1)^{i+1}\frac  {r_\mu}{r^2}\right)=0\label{eq:6.13},\ee
\be
 \sum_{i=1}^2\left(\frac 14 V_{(i)}^\nu
- \xi_{ii}^{\mu\nu}(-1)^{i+1}\frac  {r_\mu}{r^2}\right)=0.
\label{eq:6.14}\ee
In these expressions
\be
W_{(ij)}^\nu=\frac 1{2}\left(\de_i^\nu\xi^\mu_{ij\mu}+\de_{i\mu}\xi_{ij}^{\nu\mu}\right),~~~i\not = j,\ee
and
\be
V_{(i)}^\nu=\frac 1{D+2}\left(\de_i^\nu\xi_{ii\mu}^\mu+2\de_{i\mu}\xi_{ii}^{\mu\nu}\right).\ee

In the factorized case 
\be
\xi^{\mu\nu}_{ij}=\xi^{\mu}_i(x_i)\xi^\nu_j(x_j), \quad W^\nu_{ij}=\lambda_{(i)}\xi^\nu_j,
\quad V^\nu_i=2\lambda_{(i)}\xi^\nu_i,
\ee
where the $\lambda_{(i)}$'s are defined in eq. (\ref{eq:5.9}). It is easily verified that the previous equations (\ref{eq:6.11}), (\ref{eq:6.12}), (\ref{eq:6.13}) and (\ref{eq:6.14})  are satisfied. On the other hand, in principle it is possible that these equations have independent solutions, a fact that would be rather interesting.

\section{A bilocal field theory}
Bilocal field theories have been considered recently in the framework of higher spin symmetries, see for example \cite{Sorokin:2004ie}. 
These bilocal field equations are conformal invariant, therefore it is of some interest to construct a bilocal conformal invariant field theory.
This is made possible by encoding the constraint equation given in eq. (\ref{eq:3.14}) in a bilocal field theory.
To this end, let us introduce two bilocal fields, $\phi_i(x_1,x_2)$ with $i=1,2$. Then consider the action
\bea
&S=\int d^4x_1 d^4 x_2\Big[\dd{\frac 12}\left(\de_{1\mu}\phi_1(x_1,x_2)\de_1^\mu\phi_1(x_1,x_2)+
\de_{2\mu}\phi_2(x_1,x_2)\de_2^\mu\phi_2(x_1,x_2)\right)&\nn\\&-\phi_1(x_1,x_2) V(x_1,x_2)\phi_2(x_1,x_2)\Big],&\eea
where  the potential $V$ is given by
\be V(x_1-x_2)=\frac{\alpha^2}4\frac 1{(x_1-x_2)^2}.\ee
Varying with respect to $\phi_1$ and $\phi_2$ we get the equations of motion
\be
\square_1\phi_1+V\phi_2=0,~~~~\square_2\phi_2+V\phi_1=0.\label{eq:EM}\ee
Eliminating $\phi_2$ from the first equation
\be
\phi_2= -V^{-1}\square_1\phi_1,\ee
and substituting inside the second one
\be
\square_2(V^{-1}\square_1\phi_1)-V\phi_1=0.\ee
Then, multiplying by $V$ we obtain
\be
V\square_2(V^{-1}\square_1\phi_1)-V^2\phi_1=0.\label{eq:EM}\ee
Now, let us look  for solutions of the type
\be\phi_i(x_1,x_2)=e^{i(p_1 x_1 +p_2 x_2)}\tilde\phi_i(x_1-x_2).\ee
Substituting inside the equations of motion (\ref{eq:EM}) we find
\be
p_1^2\tilde\phi_1-V\tilde\phi_2=0,~~~~p_2^2\tilde\phi_2-V\tilde\phi_1=0.\ee
Eliminating again $\tilde \phi_2$ from the first one
\be
\tilde\phi_2=V^{-1}p_1^2\tilde\phi_1,\ee
and substituting inside the second one ($r=x_1-x_2$)
\be
(p_1^2p_2^2-\frac{\alpha^4}{16r^4})\tilde \phi_1(p_1,p_2,r)=0,\ee
and an analogous equation for $\tilde\phi_2$. 
Notice that what we have done here is not to take the Fourier transform of the bilocal fields, but we have simply looked at a particular solution of the field equations.

As we have seen our theory gives rise to the constraint (\ref{eq:3.14}), therefore, in quantum theory it would be natural to transform it in a wave equation of the type
 \be 
 (\square_1\square_2-\frac 1{16}\frac {\alpha^4}{r^4})\phi(x_1,x_2)=0,
 \ee
which should be looked at as a generalization the conformal invariant massless Klein-Gordon equation  to two conformal particles. On the other hand, this equation is fourth-order in the derivatives and it might produce problems in a related field theory. In fact, higher-order theories present , in general, ghosts in the spectrum. For this reason we prefer to start with a system of two fields each of them obeying  a second order equation.

 An interesting point is to expand the equations of motion  (\ref{eq:EM}) in terms of a series of higher spin local fields, but we  defer this problem to a future paper.
 
 \section{ Conformal invariant lagrangians for many particles}

In this Section we would like to extend  the case of a conformal invariant interaction between two particles to the case of $N$ particles. The kind of model one obtains depends on the symmetries one assumes in the exchange of the particles. We will start assuming the maximal symmetry, that is invariance under any permutation among the particles. This requirement and conformal invariance  fix completely the interaction among the $N$ particles up to $N-1$ dimensionless couplings.

We  start again from the einbein formulation for the lagrangian describing $N$ massless free particles
\be
L_{free}=-\sum_{i=1}^N \frac{\dot x_i^2}{2e_i}.\ee
We recall, from Section \ref{sec:2} that under inversion:
\be
\dot x_i^2\to \frac{\dot x_i^2}{x_i^4},~~~~ r_{ij}^2\to \frac{r_{ij}^2}{x_i^2 x_j^2},~~~i,j=1,2,\cdots N.\ee
In order the free part to be invariant under inversion the einbeins must transform as
\be e_i\to \frac{e_i}{x_i^4},\ee
whereas under reparametrization they must transform as a time derivative. In order to write down the invariant terms for the many particle case, let us notice that the two point interaction can be written in the following form:
\be
\left(\frac {e_ie_j}{r_{ij}^2r_{ji}^2}\right)^{1/2}.\ee
This expression suggests that an invariant term for $n$ particles is of the form
\be
\left(\frac{e_{i_1}e_{i_2}\cdots e_{i_n}}{r_{i_1i_2}^2r_{i_2i_3}^2\cdots r_{i_{n-1}i_{n}}^2
r_{i_ni_1}^2}\right)^{1/n}.\ee
In fact, it is easily seen that this term  is conformal invariant and transforms as a first derivative with respect to time.

Then, the most general conformal invariant lagrangian symmetric under the exchange of any pair of particles has the following structure
\be
L=-\sum_{i=1}^N \frac{\dot x_i^2}{2e_i}-\sum_{n=2}^N \beta_n\sum_{i_1<i_2<\cdots<i_n}\left(\frac{e_{i_1}e_{i_2}\cdots e_{i_n} }{r_{i_1i_2}^2r_{i_2i_3}^2\cdots r_{i_{n-1}i_{n}}^2
r_{i_ni_1}^2}\right)^{1/n},~~~i_k=1,2,\cdots,N. \label{eq:8.6}\ee

 It should be noticed that at difference with other theories involving $N$ particles, as for instance the Calogero model \cite{Calogero:1970nt}, in our case we have not only two-body interactions, but all the possible interactions among the $N$-particles, therefore we have a real "democratic" model. It is rather interesting that this 
arises from the requirement of conformal symmetry.

If we imagine to divide the $N$ particles in two clusters of $\{n\}$ and $\{m\}$ particles, with $n+m=N$, the previous
lagrangian can be written in the form
\be
L_N=L_n+L_m+L_{nm},\ee
where
$L_n$ and $L_m$ have exactly the same structure of $L_N$ and $L_{nm}$ contains all the terms involving distances between any  particle of the set $\{m\}$ with any particle of the set $\{n\}$. Therefore, for very large distances among the particles belonging to the two different clusters, the term $L_{nm}$ goes to zero, that is
\be
\lim_{{\rm{for ~all}}~r_{ij}^2\to\infty}{L_N}=L_n+L_m, ~~{\rm{such~that}}~i\in \{n\}, j\in\{m\}.\ee
This shows that this lagrangian is  separable for large distances among the two clusters.
We could say that our lagrangian satisfies the  cluster decomposition at the classical level.

In a complete analogous way we can get a conformal invariant lagrangian without using the einbeins. The lagrangian turns out to be
\be
L=-\sum_{n=2}^N \alpha_i\sum_{i_1<i_2<\cdots<i_n}\left(\frac{\dot x_{i_1}^2\dot x_{i_2}^2\cdots\dot x_{i_n}^2}{r_{i_1i_2}^2r_{i_2i_3}^2\cdots r_{i_{n-1}i_{n}}^2
r_{i_ni_1}^2}\right)^{1/2n}. \ee
Again this lagrangian vanishes when all the interactions are turned off.
We have not proved that this expression is obtained eliminating the einbeins  from the lagrangian (\ref{eq:8.6}), but we conjecture that this is  actually the case. The derivation of this result would be very useful in order to get  the relation between the couplings $\beta_i$ and $\alpha_i$.

As a final observation we would like to underline that the lagrangian (\ref{eq:8.6}) depends on $N-1$ dimensionless couplings $\beta_n$ and that the number of  terms  involving $n$ particles out of $N$ is given by
\be
\binom{N}{n}.\ee
Notice that the previous result applies also to the number of free terms (each of them involving one particle), by choosing $n=1$. Therefore, the total number of terms in  (\ref{eq:8.6}) is given by
\be
\sum_{n=1}^N\binom{N}{n}=2^N-1.\ee

Whereas in this model there are many particle interactions, one could consider a model with only two-body interactions. Imagine to associate the particle labels with the sites of a one-dimensional lattice and consider only nearest neighbor interactions. The action would be
\be
S=-\int d\tau \left[\sum_{i=1}^{N}\frac{\dot x_i^2}{2e_i} +\frac {\alpha^2}4 \sum_{i=1}^{N-1} \frac{\sqrt{e_i e_{i+1}}}{r_{i,i+1}^2}\right],\label{eq:213}\ee
with
\be
r_{i,i+1}=x_i-x_{i+1}.\ee
Here we have assumed a single coupling. The asymptotic cluster decomposition holds also in this case. We could also consider a closed lattice identifying the first and the last particle. In this case the model has an obvious invariance under discrete translations. 
Furthermore, it can be extended to the continuum obtaining a conformal string (open or closed). This extension is actually under study and it will be the object of a different publication \cite{work}.

\section{Conclusions and outlook}

In this paper we have studied what relativistic conformal symmetry can teach us about  possible interactions among $N$ classical massless particles. The lagrangian considered here depends on the symmetry we assume under the exchange of the particles. Assuming 
 invariance under any permutation among the particles, the lagrangian is completely fixed up to $N-1$ dimensionless coupling constants. This lagrangian is rather interesting because it does not contain  two-body interactions only, but for any subset of $n$ particles out of the total set, it contains $n$-body interactions and it appears to be completely democratic.  This is also shown by the number of terms in the lagrangian, which does not grow with a power of $N$ but rather in an exponential way, namely like {\bf $2^N-1$. } Another possibility that we have considered is the one corresponding to nearest neighbor interactions. The interest of this case is mainly related to the possibility of getting a simple limit in the continuum, obtaining in this way a conformal string  
\cite{work}. 

We have analyzed  the case $N=2$ with a particular emphasis on the symmetries. In fact, it is known that the conformal symmetry of the free massless Klein-Gordon equation can be extended to the enveloping algebra of the conformal group, obtaining in this way higher spin symmetries. In our case, we have two (or more) interacting particles preserving conformal symmetry, so a natural question to investigate is the possibility to enlarge the higher spin symmetries to interacting massless particles.

An interesting point is  the extension of the conformal models presented here to superconformal ones \cite{work}.

The $N$ particle models could be considered in the case of $D=1$, that is a pure quantum mechanical case, in order to study their possible integrability. In particular, the nearest neighbor model looks close to the Calogero-type models \cite{Calogero:1970nt,Moser:1975qp,Peremelov:1983aa}.

Another  problem, to be investigated in the future, is the  quantization of these models. A promising possibility, in our opinion, would be to try the world-line quantization, along the way paved by string theory. We recall here that the world-line quantization can be extended to the self-interactions of scalar particles \cite{Casalbuoni:1974pj}.  Another option is  field quantization using, in the case of two particles, bilocal fields as introduced in Section 7.

\acknowledgements

We thank Eric Berghsoeff and Jaume Gomis for their helpful comments. One of us (J.G.) acknowledges  the hospitality at the Department of Theoretical Physics of the 
University of Groningen, where this work was started. This work is partially financed by the Dutch research organization   FOM
and  FPA 2010-20807, 2009 SGR502, CPAN, Consolider CSD 2007-0042.

\end{document}